\documentclass[prd,preprint,superscriptaddress,amsmath,amssymb]{revtex4}
\usepackage{graphicx,color,slashed}

\usepackage{subfigure,bbold}
\usepackage[toc,page]{appendix}

\def\be{\begin{eqnarray}}
\def\ed{\end{eqnarray}}
\def\non{\nonumber}

\def\lam{\lambda}

\begin{document}

{\begin{flushright}{KIAS-P15060}
\end{flushright}}

\title{ $h, Z\to \ell_i \bar\ell_j$, $\Delta a_{\mu}$, $\tau\to (3\mu,\mu \gamma)$  in generic two-Higgs-doublet models }

\author{  R.~Benbrik \footnote{Email: rbenbrik@ictp.it}}
\affiliation{LPHEA, Semlalia, Cadi Ayyad University, Marrakech, Morocco}
\affiliation{MSISM Team, Facult\'e Polydisciplinaire de Safi, Sidi Bouzid B.P 4162, 46000 Safi, Morocco}

\author{ Chuan-Hung Chen \footnote{Email: physchen@mail.ncku.edu.tw} }
\affiliation{Department of Physics, National Cheng-Kung University, Tainan 70101, Taiwan }

\author{ Takaaki Nomura \footnote{Email: nomura@kias.re.kr }}
\affiliation{School of Physics, Korea Institute for Advanced Study, Seoul 130-722, Republic of Korea}

\date{\today}

\begin{abstract}
Inspired by a  significance of $2.4\sigma$ for the $h\to \mu \tau$ decay observed  by the CMS experiment at $\sqrt{s} = 8$ TeV, we investigate the Higgs lepton-flavor-violating effects in a generic two-Higgs-doublet model (THDM), where the lepton-flavor-changing neutral currents are induced at the tree level and arise from Yukawa sector. We revisit the constraints for generic THDM by considering theoretical requirements, precision measurements of $\delta \rho$ and oblique parameters $S$, $T$, and $U$, and  Higgs measurements. The bounds from Higgs data play the major limits. With  parameter values that  simultaneously satisfy the Higgs bounds and the CMS excess of the Higgs coupling to $\mu$-$\tau$, we find that the tree-level $\tau\to 3\mu$ and the  loop-induced $\tau\to \mu \gamma$ decays are  consistent with current experimental upper limits; the discrepancy in muon $g-2$ between experimental results and  standard model predictions can be resolved, and  an interesting relation between muon $g-2$ and the branching ratio (BR) for $\mu\to e \gamma$ is found. The generic THDM results show that the order of magnitude of the ratio $BR(h\to e\tau)/BR(h\to \mu \tau)$ is smaller than  $10^{-4}$. Additionally, we also study the rare decay $Z\to \mu \tau$ and get $BR(Z\to \mu\tau)< 10^{-6}$.  

\end{abstract}

\maketitle

   The observed-flavor-changing neutral currents (FCNCs)   in the standard model (SM) occur at the loop level of the quark sector and originate from $W$-mediated charged currents, such as $K-\bar K$, $B-\bar B$, and $D-\bar D$ mixings and $b\to s\gamma$.  Due to the loop effects, it is believed that these FCNC processes are sensitive to new physics. However, most of these processes involve large uncertain non-perturbative quantum chromodynamics (QCD) effects; therefore, even if new physics exist, it is not easy to distinguish them from the SM results due to QCD uncertainty.
 
The situation in the lepton sector is different. Although the SM also has lepton FCNCs (e.g., $\mu\to e \gamma$ and $\tau \to (e,\mu)\gamma$)  they are irrelevant to QCD effects and highly suppressed; if any signal is observed, it is certainly strong evidence for new physics. It is thus  important to search for new physics through the lepton sector~\cite{Crivellin:2013wna,Gomez:2014uha,Crivellin:2015hha}. 

With the discovery  of a new  scalar with  a mass of around 125 GeV at the ATLAS~\cite{:2012gk} and CMS~\cite{:2012gu} experiments, we have taken one step further  toward understanding the electroweak symmetry breaking (EWSB) through spontaneous symmetry breaking (SSB) mechanism in the scalar sector. With  $\sqrt{s}=13 - 14$ TeV, the next step for the High Luminosity Large Hadron Collider (LHC) is to explore not only the detailed properties of the observed scalar, but also  the existence of other Higgs scalars and  new physics effects.   

CMS~\cite{Khachatryan:2015kon} and ATLAS~\cite{Aad:2015gha}  have recently reported the measurements of $h\to \mu \tau$ decay in $pp$ collisions at $\sqrt{s}=8$ TeV. At the $95\%$ confidence level (CL),  the branching ratio (BR) for the decay at CMS is $BR(h\to \mu\tau)<1.51\%$ and that at 
ATLAS is $BR(h\to \mu\tau)<1.85\%$ . Additionally, a slight excess of events with  a significance of $2.4\sigma$ was reported by CMS, with the best fit of $BR(h\to \mu\tau)=(0.84 ^{+0.39}_{-0.37})\%$. If the excess is not a statistical fluctuation, the extension of the SM becomes necessary. Inspired by the excess of events, possible new physics effects have been studied~\cite{Campos:2014zaa,Sierra:2014nqa,Lee:2014rba, Heeck:2014qea,Crivellin:2015mga,Dorsner:2015mja,Omura:2015nja,Crivellin:2015lwa,Das:2015zwa,Bishara:2015cha,Varzielas:2015joa,He:2015rqa,Chiang:2015cba,Altmannshofer:2015esa,Cheung:2015yga,Arganda:2015naa,Botella:2015hoa,Baek:2015mea,Huang:2015vpt,Baek:2015fma,Arganda:2015uca,Aloni:2015wvn}. 
The earlier relevant works  have been investigated~\cite{Arganda:2004bz, Blankenburg:2012ex,Arhrib:2012mg,Harnik:2012pb,Dery:2013rta,Arana-Catania:2013xma,Arroyo:2013tna,Celis:2013xja,Falkowski:2013jya,Arganda:2014dta,Dery:2014kxa}.

Following the measurements of ATLAS and CMS of the couplings of Higgs to leptons, we  investigate the lepton flavor violation (LFV) in a generic two-Higgs-doublet  model (THDM)~\cite{Lee:1973iz}. The THDM includes  five physical scalar particles, namely  two CP-even bosons, one CP-odd pseudoscalar, and one charged Higgs boson.  According to the  form in which Higgs doublets couple to fermions, the THDM is classified  as type I, II, and III models,  lepton-specific model, and flipped model~\cite{Branco:2011iw}. The minimal supersymmetric SM (MSSM) belongs to the type II THDM, in which one Higgs doublet couples to up-type quarks while the other  couples to down-type quarks. The type III THDM corresponds to the case in which   each of the two Higgs doublets couples to all fermions simultaneously. As a result, tree-level FCNCs in the quark and charged lepton sectors are induced. Considering the strict  experimental data, it is interesting to determine the impacts of the type III model on  the LFV. 

  If we assume no new CP-violating source from the scalar sector, such as the  type II model and MSSM,  the main new free parameters are the masses of new scalars, $\tan\beta=v_2/v_1$ and angle $\alpha$, where $\tan\beta$ is related to the ratio of the vacuum expectation values (VEVs) of two Higgs fields and the angle $\alpha$ stands for the mixing effect of two CP-even scalars. Basically, these two parameters have been strictly constrained by the current experimental data, such as  $\rho$-parameter, $S$, $T$, and  $U$ oblique parameters, Higgs searches through $h\to (\gamma\gamma, WW^*, ZZ^*, \tau\tau, b\bar b)$, etc. In order to show the correlation of free parameters and these experimental bounds, we revisit the constraints by adopting the $\chi$-square fitting approach. it can be seen that although the allowed values of $\cos(\beta-\alpha)$ approach the decoupling limit (i.e., $\alpha\sim \beta- \pi/2$)  if $\cos\beta$ is sufficiently small, the BR for $h\to \mu \tau$ could still be as large as the measurements from ATLAS and CMS. 
 
Besides the $h\to \ell_i \bar\ell_j$ decays, the type III model has also significant effects on other lepton-flavor-conserving and -violating processes, such as muon anomalous magnetic moment, $\mu\to 3 e$, $\mu (\tau) \to e (\mu, e) \gamma$, $Z\to \ell_i \bar\ell_j$, etc. Although concrete signals for lepton-flavor-violating processes have not been observed yet, the current experimental data with $BR(\mu\to 3 e)< 10^{-12}$ and $BR(\mu\to e \gamma)<5.7\times 10^{-13}$~\cite{PDG} have put strict limits on  $\mu \to 3 e$ and $\mu\to e \gamma$, respectively. Combing the LHC data  and the upper limits  of the rare lepton decays, we study whether the excess of muon $g-2$ can be resolved and whether the BRs of the listed lepton FCNC processes  are consistent with current data   in the type III THDM. 

To indicate the scalar couplings to fermions in the type III model, we express  the Yukawa sector as:
\be
-{\cal L}_Y &=& \bar Q_L Y^u_1 U_R \tilde H_1 + \bar Q_L Y^{u}_2 U_R
\tilde H_2  \non \\
&+& \bar Q_L Y^d_1 D_R H_1 + \bar Q_L Y^{d}_2 D_R H_2 \non \\
&+&  \bar L Y^\ell_1 \ell_R H_1 + \bar L Y^{\ell}_2 \ell_R H_2 + h.c.\,, 
\label{eq:Yu}
\ed
where we have hidden all flavor indices, $Q^T_L=(u, d)_L$ and $L^T = (\nu, \ell)_L$ are the $SU(2)_L$ quark and lepton doublets, respectively, $Y^f_{1,2}$ are the Yukawa matrices, $\tilde H_i = i\tau_2 H^*_i$ with $\tau_2$ being the second Pauli matrix, the Higgs doublets are represented by:
\be
H_i &=& \left(
            \begin{array}{c}
              \phi^+_i \\
              (v_i+\phi_i +i \eta_i)/\sqrt{2} \\
            \end{array}
          \right)\,, \label{eq:doublet}
\ed
and $v_i$ is the VEV of $H_i$. Equation~(\ref{eq:Yu}) can  recover  the type II THDM  if $Y^u_1$, $Y^d_2$, and $Y^\ell_2$ vanish. Before EWSB, all $Y^{f}_{1,2}$ are arbitrary $3\times 3$ matrices and fermions are not physical eigenstates; therefore, we have the freedom to choose $Y^u_1$, $Y^d_2$, and $Y^\ell_2$ to have diagonal forms; that is, $Y^u_1={\rm diag}(y^u_1, y^u_2, y^u_3)$ and 
$Y^{d,\ell}_{2}= {\rm diag}(y^{d,\ell}_1, y^{d,\ell}_2, y^{d\ell}_{3})$.

By the measurements of  neutrino oscillations, it is known that  the SM neutrinos are massive particles. Since the origin of neutrino masses is not conclusive, in order to introduce the neutrino masses and avoid changing the structure of scalar interactions in the THDM, the neutrino masses can be generated through the type-I seesaw mechanism~\cite{T1seesaw}. If we include three heavy right-handed neutrinos, the associated Yukawa couplings are given by:
 \be
 -Y_{N} = \left( \bar L \left({\bf y}_1  \tilde H_1 + {\bf y}_2 \tilde H_2 \right) N + h.c.\right)+ \frac{1}{2} {\cal N}^T C {\bf m}_{N} {\cal N}  \,,
 \ed
where we have suppressed the flvaor indices, ${\cal N}_i =N_i$ stands for the heavy right-handed neutrino, and diag${\bf m}_N=(m_{N1}, m_{N2}, m_{N3})$ in flavor space. Accordingly, the neutrino mass matrix is expressed as:
 \be
 m_\nu = \left(  \begin{array}{cc}
  \mathbb{0} & ({\bf y}_1 v_1 + {\bf y}_2 v_2)/\sqrt{2} \\
   ({\bf y}^T_1 v_1 + {\bf y}^T_2 v_2)/\sqrt{2} & {\bf m}_N
 \end{array} \right)\,.
 \ed
By taking proper values of ${\bf m}_N$ and ${\bf y}_{1,2}$, we can fit the the measured mass-square differences, where the data are $\Delta m^2_{12} =(7.53\pm 0.18)\times 10^{-5}$ eV$^2$ and $\Delta m^2_{23} = (2.44\pm 0.06)\times 10^(-3)$ eV$^2$ for normal hierarchy, or $m^2_{23} =(1.52 \pm 0.07)\times 10^{-3}$ eV$^2$ for inverted hierarchy~\cite{PDG}. Since the neutrino effects are irrelevant to the current study,  hereafter we do not further discuss the detailed properties of neutrinos.

The VEVs $v_{1,2}$ are dictated by the scalar potential, where  the gauge invariant form is given by \cite{Branco:2011iw}:
 \be
V(\Phi_1,\Phi_2)
&=& m^2_1 \Phi^{\dagger}_1\Phi_1+m^2_2 \Phi^{\dagger}_2\Phi_2 -(m^2_{12}
\Phi^{\dagger}_1\Phi_2+{\rm h.c}) +\frac{1}{2} \lam_1 (\Phi^{\dagger}_1\Phi_1)^2 
\nonumber \\ &+& \frac{1}{2} \lam_2
(\Phi^{\dagger}_2\Phi_2)^2 +
\lam_3 (\Phi^{\dagger}_1\Phi_1)(\Phi^{\dagger}_2\Phi_2) + \lam_4
(\Phi^{\dagger}_1\Phi_2)(\Phi^{\dagger}_2\Phi_1)  \non \\
&+& 
 \left[\frac{\lam_5}{2}(\Phi^{\dagger}_1\Phi_2)^2 +  \left(\lam_6 \Phi^\dagger_1 \Phi_1 + \lam_7 \Phi^\dagger_2 \Phi_2 \right) \Phi^\dagger_1 \Phi_2+{\rm h.c.} \right]\,.
\label{eq:higgspot}
 \ed
Since we do not concentrate on  CP violation, we set the parameters in Eq.~(\ref{eq:higgspot}) to be real numbers. In addition, we also require the CP phase that arises from the ground state to vanish~\cite{Lee:1973iz}. By the  scalar potential with CP invariance, we have nine free parameters. In our approach, the independent parameters are taken as:
 \be
\{ m_{h}\,, m_{H}\,, m_{A}\,, m_{H^\pm}\,, v\,, \tan\beta\,,  \alpha\,, \lambda_6\,, \lambda_7 \}
\label{eq:parameters} 
\ed
with $v=\sqrt{v^2_1+ v^2_2}$. 

With the nonvanished $\lambda_{6,7}$ terms in the potential, not only the  mass relations of 
scalar bosons are modified but also the scalar triple and quartic couplings 
receive the changes. Since the masses of scalar bosons are treated as free parameters,  the direct contributions of  $\lambda_{6,7}$  to the process $h\to \gamma\gamma$ in this study are  through  the triple coupling $h$-$H^+$-$H^-$. By the constraint from $b\to s\gamma$, the mass of charged Higgs can not be lighter than 480 GeV;  the contribution of the charged Higgs loop  to the decay $h\to \gamma\gamma$  is small. That is,  the influence of  $\lambda_{6,7}$ on the constraint of parameter is not significant. Without loss of generality, in the phenomenological analysis, we set $\lambda_{6,7}\ll 1$. The detailed numerical study with $\lambda_{6,7}\sim {\cal O}(1)$ can be found elsewhere~\cite{Arhrib:2015maa}. 

The physical states for scalars are expressed by: 
\be
 h &=& -s_\alpha \phi_1 + c_\alpha  \phi_2 \,, \non \\
 H&=& c_\alpha \phi_1 + s_\alpha \phi_2 \,, \non \\
H^\pm (A) &=& -s_\beta \phi^\pm_1 (\eta_1) + c_\beta \phi^\pm_2 (\eta_2) \label{eq:hHHA}
\ed
with $c_\alpha (s_\alpha)= \cos\alpha (\sin\alpha)$, $c_\beta  = \cos\beta = v_1/v$, and $s_\beta= \sin\beta = v_2/v$. In this study, $h$ is the SM-like Higgs while $H$, $A$, and $H^\pm$ are new particles in the THDM. 
Using Eqs.~(\ref{eq:Yu}) and (\ref{eq:doublet}), one can easily find that the fermion mass matrix  is 
\be
{\bf  M}_f =  \frac{v}{\sqrt{2}}\left( \cos\beta  Y^f_1 + \sin\beta  Y^f_2 \right)\,.
\ed
If we introduce the unitary matrices $V^f_L$ and $V^f_R$, the mass matrix can be diagonalized through ${\bf m}_f = V^f_L {\bf M}_f V^{f\dagger}_R$. Accordingly, the scalar  couplings to fermions could be formulated as:
\be
-{\cal L}_{Y\phi} &=&  \bar \ell_L \; \epsilon_\phi {\bf y}^\ell_{\phi} \; \ell_R\; \phi +  \bar \nu_L V_{\rm PMNS} \;  {\bf y}^\ell_{H^\pm} \; \ell_R\; H^+ + h.c. \,,\label{eq:Yphi}
\ed
where $\phi=h, H, A$ stands for the possible neutral scalar boson, $\epsilon_{h(H)}=1$, $\epsilon_A= i $, $V_{\rm PMNS}$ is the Pontecorvo-Maki-Nakagawa-Sakata matrix, and the Yukawa couplings ${\bf y}^\ell_{\phi, H^\pm}$ are defined by:
 \begin{align}
 ({\bf y}^\ell_h)_{ij} &= -\frac{s_\alpha}{c_\beta} \frac{m_i}{v} \delta_{ij} + \frac{c_{\beta\alpha}}{ c_\beta} X^\ell_{ij}\,, \non  \\
 ({\bf y}^\ell_H)_{ij} &= \frac{c_\alpha}{c_\beta} \frac{m_i}{v} \delta_{ij} - \frac{s_{\beta\alpha}}{ c_\beta} X^\ell_{ij}\,, \non \\
 ({\bf y}^\ell_A)_{ij} &= - \tan\beta \frac{m_i}{v} \delta_{ij} + \frac{ X^\ell_{ij}}{ c_\beta} \,, \label{eq:YhHA}
 \end{align}
and $ {\bf y}^\ell_{H^\pm} =\sqrt{2} {\bf y}^\ell_{A}$ with $c_{\beta\alpha} = \cos(\beta-\alpha)$, $s_{\beta\alpha}= \sin(\beta-\alpha)$ and 
 \be
  {\bf X^u}= V^u_L \frac{Y^u_1}{\sqrt{2}}  V^{u\dagger}_R \,, \     {\bf X^d}= V^d_L \frac{Y^d_2}{\sqrt{2}} V^{d\dagger}_R \,, \ 
{\bf X^\ell}= V^\ell_L \frac{Y^\ell_2}{\sqrt{2}} V^{\ell\dagger}_R\,. \label{eq:Xs}
 \ed
 From these  formulations, it can be seen that the Yukawa couplings of Higgses to fermions can return to  the type II THDM when $Y^u_1$ and $Y^{d,\ell}_2$ vanish. The FCNC effects are also associated with $Y^u_1$ and $Y^{d,\ell}_{2}$, which can be chosen to be diagonal matrices, as mentioned earlier. The detailed Yukawa couplings of $H$, $A$, and $H^\pm$ to up- and down-type quarks are summarized in the appendix. 

In principle, $Y^f_{1,2}$ are arbitrary free parameters. In order to get more connections among parameters and reduce the number of free parameters, the hermitian Yukawa matrices can be applied, where the hermiticity of the Yukawa matrix can be realized  by symmetry, such as global (gauged) horizontal $SU(3)_H$ symmetry~\cite{Masiero:1998yi} and  left-right symmetry~\cite{Babu:1999xf}. Therefore,  the equality $V^f_L = V^f_R \equiv V^f$ can be satisfied. With the diagonal $Y^u_1$ and $Y^{d,\ell}_2$, the $Xs'$ effects in Eq.~(\ref{eq:Xs}) can be expressed as $
 X^f_{ij} = V^f_{ik} V^{f*}_{jk} y^f_k$, where the index $k$ is summed up. Since no CP violation is observed in the lepton sector, it is reasonable to assume that $Y^\ell_{1,2}$ are real numbers. Based on this assumption,  ${\bf X}^\ell$ is a symmetric matrix, i.e., $X^\ell_{ij} = X^\ell_{ji}$.  In the decoupling limit of $\alpha=\beta-\pi/2$,  the Yukawa couplings in Eq.~(\ref{eq:YhHA})  become:
  \begin{align}
 ({\bf y}^\ell_h)_{ij} &= \frac{m_i}{v} \delta_{ij} \,, \non  \\
 ({\bf y}^\ell_H)_{ij} &= -({\bf y}^\ell_A)_{ij} = \tan\beta \frac{m_i}{v} \delta_{ij} - \frac{1}{ c_\beta} X^\ell_{ij}\,.
 %
 \end{align}
In such a limit, we see that the tree-level lepton FCNCs are suppressed in $h$ decays; however, they are still allowed in $H$ and $A$ decays. 

Next, we discuss the scalar-mediated lepton-flavor-violating effects on the processes of interest. Using the couplings in Eq.~(\ref{eq:Yphi}),  the BR for $h \to \tau \mu$  is given by:
\be
BR(h \to \mu \tau) = \frac{c_{\beta \alpha}^2 (|X^\ell_{23}|^2 + |X^\ell_{32}|^2)}{16 \pi c_\beta^2 \Gamma_h} m_h. \label{eq:brhmutau}
\ed
With $m_h=125$ GeV, $\Gamma_h\approx 4.21$ MeV, and $X^\ell_{32} = X^\ell_{23} $, we can express $X^{\ell}_{23}$ as
 \be
  X^\ell_{23} = 3.77\times 10^{-3}\left(\frac{c_{\beta}}{0.02}\right) \left(\frac{0.01}{c_{\beta\alpha}}\right) \sqrt{\frac{BR(h\to \mu \tau )}{0.84\times 10^{-2}}}\,, 
 \ed
where $BR(h\to \mu \tau)$ can be taken from the experimental data. If one adopts the ansatz $ X^\ell_{\mu\tau}=\sqrt{m_\mu m_\tau}/v \chi^\ell_{\mu\tau}$, $\chi^\ell_{\mu\tau}\sim 2$ fits the current CMS excess. 

 Moreover, we find that the same $X^\ell_{23}$ effects also contribute to  the decay $\tau \to 3\mu$  at tree level through the mediation of scalar bosons. The BR can be formulated as:
\be
BR(\tau \to 3 \mu) = \frac{  \tau_\tau m^5_\tau }{3 \cdot 2^9 \pi^3} \frac{|X^\ell_{23}|^2}{c^2_\beta}\left[ \left| \frac{c_{\beta\alpha}y^\ell_{h22} }{m^2_h} -\frac{s_{\beta\alpha} y^\ell_{H22}}{m^2_H}\right|^2 + \left| \frac{  y^\ell_{A 22}}{m^2_A}\right|^2\right] \label{eq:tau3mu}
\ed
with $\tau_\tau$ being the lifetime of a tauon.  Equation~(\ref{eq:tau3mu}) can be applied to $\mu \to 3e$ when the corresponding quantities are correctly replaced. If we set $X^\ell_{ij} = \sqrt{m_i m_j}/v \chi^\ell_{ij}$ and assume that $\chi^\ell_{ij} =\chi^\ell$ are independent of lepton flavors, the ratio of $BR(\mu\to 3e)$ to $BR(\tau\to 3\mu)$  can be naively estimated as:
 \be
 R_{\mu/\tau} \sim \frac{\tau_\mu}{\tau_\tau} \frac{m^5_\mu}{m^5_\tau} \frac{m^3_e }{m_\tau m^2_\mu} = 3.5\times 10^{-8}. \label{eq:Rmu/tau}
 \ed
 With the current upper limit $BR(\tau \to 3 \mu) < 1.2 \times 10^{-8}$~\cite{Amhis:2014hma}, we get  $BR(\mu\to 3 e)< 4.2 \times 10^{-16}$  in the type III model, which is far smaller than the current upper bound. Nevertheless, the suppression factor of $m^3_e/(m_\tau m^2_\mu)$ in Eq.~(\ref{eq:Rmu/tau}) can be relaxed to be $m_e/m_\tau$ at the one-loop level, where the lepton pair is  produced by virtual $\gamma/Z$ in the SM.
 Since the $X^\ell_{23}$ parameter  also appears in the decays  $\mu\to e \gamma$ and $\tau\to \mu \gamma$, which have stronger limits in experiments, in the following analysis we do not further discuss these processes. Additionally, to remove the correlation between $\tau \to  3\mu$ and $\mu\to 3 e$, $\chi^\ell_{ij}$  should  be taken as being flavor-dependent.  

The discrepancy in muon $g-2$ between experimental data and the SM prediction now is $\Delta a_\mu = a^{\rm exp}_\mu - a^{\rm SM}_\mu =(28.8\pm 8.0)\times 10^{-10}$~\cite{PDG}. Although muon $g-2$ is a flavor-conserving process, $X^{\ell}_{23}$ and $X^\ell_{21}$ also contribute to the anomaly through loops that are mediated by neutral and charged Higgses. Thus, 
the muon anomaly  in the type III model can be formulated as~\cite{Assamagan:2002kf,Davidson:2010xv}:
\be
\Delta a_\mu &\simeq& \frac{m_\mu m_\tau X^{\ell}_{23} X^{\ell}_{32}}{8 \pi^2 c_\beta^2} Z_\phi \,, \label{eq:g-2} \\
Z_\phi &=&  \frac{c_{\beta \alpha}^2 \left( \ln( m_h^2/m_\tau^2 )-\frac{3}{2} \right)}{m_h^2} 
 +  \frac{s_{\beta \alpha}^2 \left( \ln (m_H^2/m_\tau^2) -\frac{3}{2} \right)}{m_H^2} \non \\
&-&  \frac{  \ln (m_A^2/m_\tau^2 )- \frac{3}{2} }{m_A^2}\,, \non 
\ed
where we have dropped the subleading terms associated with $m^2_\mu$.   The following question is explored below: when the current strict experimental data are considered, can the anomaly of $\Delta a_\mu$  be explained in the type III model? 


As mentioned earlier, the radiative lepton decays $\mu\to e\gamma$ and $\tau\to (\mu, e) \gamma$ in the SM are very tiny and sensitive to new physics effects. In the type III model, these radiative decays can be generated by charged and neutral Higgses through the FCNC effects. For illustration, we present the following effective interaction for  $\mu\to e \gamma$:
\be
{\cal L}_{\mu \to e \gamma} =\frac{ e m_\mu }{16\pi^2}  \bar e \sigma_{\mu \nu} \left( C_L P_L + C_R P_R\right) \mu  F^{\mu \nu}\,,
\ed
where $F^{\mu\nu}$ is the electromagnetic field strength tensor, and the Wilson coefficients $C_L$ and $C_R$ from neutral and charged scalars are given by:
\begin{align}
C_{L(R)}& = C^{ \phi}_{L(R)}  + C^{ H^\pm}_{L(R)}\,, \non \\
C^{ \phi}_L & = \frac{ X^\ell_{32} X^\ell_{13} }{2 c^2_\beta} \frac{m_\tau }{m_\mu} Z_\phi \,, \non \\ 
C^{ H^\pm}_L &= - \frac{1}{12 m^2_{H^\pm}}  \left(\frac{2 X^{\ell}_{23} X^\ell_{13}}{c^2_\beta}\right)\,, \label{eq:muWCs}
\end{align}
where $C^{ \phi}_{R} = C^{ \phi}_{L}$, $C^{ H^\pm}_R=0$, and  the BR for $\mu\to e \gamma$ is:
 \be
 \frac{BR(\mu \to e \gamma)}{BR(\mu \to e \bar \nu_e \nu_\mu) } = \frac{3 \alpha_e}{4\pi G^2_F} \left( |C_L|^2 + |C_R|^2\right)\,.
 \ed
It is clear that the factor  $Z_\phi$  in $\Delta a_\mu$ also appears in $C^\phi_{L(R)}$. In terms of $\Delta a_\mu$ in Eq. (\ref{eq:g-2}),  $C^\phi_{L(R)}$ can be expressed as:
 \be
 C^\phi_{L(R)} = \frac{X^\ell_{13}}{X^\ell_{23}} \frac{4\pi^2 \Delta a_\mu }{m^2_\mu}\,. \label{eq:CphiL}
 \ed
 Since $C^\phi_{L(R)}$ has an enhancement  factor of $m_\tau/m_\mu$, the contribution from charged Higgs becomes the subleading effect. The formulas for $\tau \to \mu \gamma$ can be found in the appendix.  From Eq.~(\ref{eq:muWCs}), we see that if flavor-changing effects $X^{\ell}_{ij} =0$ with $i\neq j$,  the effective Wilson coefficients $C_{L, R}$ vanish. That is, the contributions  to the radiative lepton decays from other types of THDM are suppressed. Therefore, any sizable signals of $\mu\to e \gamma$ and $\tau\to \mu \gamma$ will be a strong support for the type III model. 

\begin{figure}[!ht] 
\includegraphics*[width=120mm]{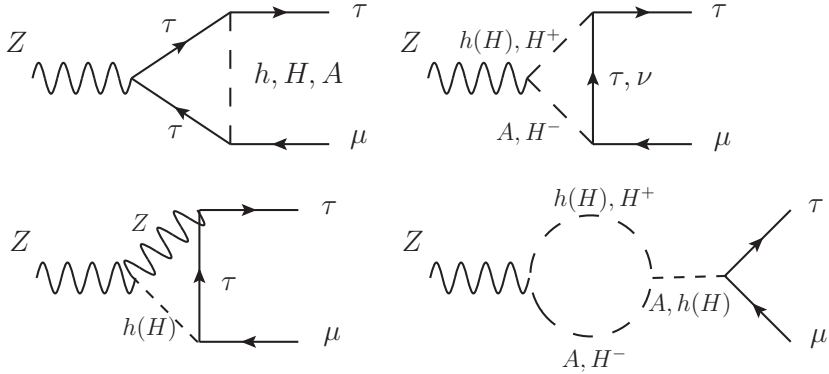} 
\caption{ Representative Feynman diagrams for $Z\to \mu \tau$ decay. }
\label{fig:Ztaumu}
\end{figure}

The last process of interest  is the decay $Z\to \mu \tau$. Other flavor-changing leptonic $Z$ decays also occur in the type III model; however, since the $\mu \tau$ mode is dominant, the present study  focuses on the $\mu \tau$ channel. Besides the $Z$ coupling  to charged leptons, in the THDM,  $Z$-$h(H)$-$A$ and $Z$-$Z$-$h(H)$ interactions are involved, in which the vertices are \cite{Gunion:1989we}:
 \begin{align}
 Z-h(H)-A& : ~~~ -\frac{g c_{\beta\alpha} ( -s_{\beta\alpha})}{2\cos\theta_W} ( p_A + p_{h(H)})_\mu \,, \non \\
 Z-H^+-H^- &: ~~~ -i \frac{g\cos2\theta_W}{2\cos\theta_W}( p_{H^+}+ p_{H^-})_\mu\,, \non \\
 Z-Z-h(H) &:~~~  \frac{g m_Z}{\cos\theta_W}s_{\beta\alpha} (c_{\beta\alpha}) g_{\mu\nu}
 \end{align}
where $\theta_W$ is  Weinberg's angle. The typical Feynman diagrams for $Z\to \mu \tau$ are presented in Fig.~\ref{fig:Ztaumu}.
 Since many one-loop Feynman diagrams are involved  in the process, we employ the FormCalc  package~\cite{Hahn:1998yk} to deal with the loop calculations. The lengthy formulas are not shown here; instead, we directly show the numerical results. 
  
 Before presenting  the numerical analysis, we discuss the theoretical and experimental constraints. The main theoretical  constraints of THDM are the perturbative scalar potential, vacuum stability, and unitarity. Therefore, in order to satisfy the perturbative requirement, we set  all quartic couplings of the scalar potential to obey $|\lambda_i| \leq 8 \pi$ for all $i$. 
 The conditions for vacuum stability are~\cite{Sher:1988mj,Ferreira:2004yd}:
\begin{eqnarray}
\nonumber
&& \lambda_1  > 0\;,\quad\quad \lambda_2 > 0\;, \lambda_3 +
  \sqrt{\lambda_1\lambda_2 } > 0, \quad 
 \sqrt{\lambda_1\lambda_2 }
+ \lambda_{3} + \lambda_{4}  -|\lambda_{5}| >0,\\ &&
{2|\lambda_6 + \lambda_7| \le \frac{1}{2}(\lambda_1 + \lambda_2) +
  \lambda_3 + \lambda_4 + \lambda_5}\,.
\label{vac}
\end{eqnarray}
Without losing the general properties, we set $\lambda_{6,7}\ll 1$ in our numerical analysis. Effectively, the scalar potential is similar to that in the type II THDM. Since the unitarity constraint involves a variety of scattering processes,  here we adopt the results of a previous study~\cite{Akeroyd:2000wc}.

Next, we briefly state the experimental bounds. 
It is known that $b\to s \gamma$ is sensitive to the mass of charged Higgs. 
According to a recent analysis~\cite{Misiak:2015xwa}, the lower bound in the type II model is given to be $m_{H^\pm} > 480$ GeV at 95$\%$ CL.  Due to the neutral and charged Higgses involved  in the self-energy of W and Z bosons,   the precision measurements of the $\rho$-parameter and the oblique parameters~\cite{Peskin:1991sw} can give  constraints on  the associated new parameters. From the global fit,  we know that $\rho= 1.00040\pm 0.00024$~\cite{PDG} and  the SM prediction is $\rho=1$. Taking $m_h = 125$ GeV,  $m_t = 173.3$ GeV, and assuming  $U=0$,  the tolerated ranges for S and T are found to be~\cite{Baak:2014ora}:
\begin{eqnarray}
 \Delta S = 0.06\pm0.09\,, \ \  \Delta T = 0.10\pm0.07\,,
\label{test:ST}
\end{eqnarray}
where the correlation factor is $\rho=+0.91$, $\Delta S = S^{\textrm{2HDM}} - S^{\textrm{SM}}$, 
$\Delta T = T^{\textrm{2HDM}} - T^{\textrm{SM}}$, and their explicit expressions  can be found elsewhere~{\cite{Gunion:2002zf}}. We note that in the limit $m_{H^\pm}=m_{A}$ or $m_{H^\pm}=m_{H}$,
$\Delta T$ vanishes~\cite{Gerard:2007kn,Cervero:2012cx}. 

Since the  Higgs data approach the precision measurements, the relevant measurements can give  strict limits on $c_{\beta \alpha}$ and $s_\alpha$. As usual,  the Higgs measurement is expressed by the signal strength, which is defined by the ratio of the Higgs signal to the SM prediction, given by:
\begin{eqnarray}
\mu^f_i= \frac{\sigma_i(h) \cdot BR(h\to f) }{\sigma^{SM}_i(h)
\cdot BR^{SM}(h\to f) } \equiv \bar \sigma_i \cdot \mu_f \,. \label{eq:kvf}
\end{eqnarray}
$\sigma_i(h)$ denotes the Higgs production cross section by channel $i$
and $BR(h\to f)$ is the BR for the Higgs decay $h\to
f$. Since several Higgs boson production channels are available at the LHC, we
are interested in the gluon fusion production ($ggF$), $t \bar t h$, vector
boson fusion (VBF) and Higgs-strahlung $Vh$ with $V=W/Z$; they are grouped 
to be $\mu^f_{ggF+t\bar t h}$ and $\mu^f_{VBF+Vh}$. The values of observed signal strengths  are shown in Table.~\ref{tab:Higgs data},  where  we used the notations $\widehat{\mu}^f_{ggF+t\bar t h}$ and
$\widehat{\mu}^f_{VBF+Vh}$ to express the combined results of ATLAS~\cite{atlas034}
and CMS~\cite{cms005}. 
\begin{table}[!ht]  
\caption{Combined best-fit signal strengths $\widehat{\mu}_{\rm{ggF+tth}}$ and 
$\widehat{\mu}_{\rm{VBF+Vh}}$ 
and the associated correlation coefficient $\rho$ for corresponding Higgs decay 
 mode~\cite{atlas034,cms005}.}
\begin{ruledtabular}
\begin{tabular}{c|ccccc}
$f$ & $\widehat{\mu}^{f}_{\rm{ggF+tth}}$ & $\widehat{\mu}^{f}_{\rm{VBF+Vh}}$ &$\pm\,\,1\widehat{\sigma}_{\rm{ggF+tth}}$
&  $\pm\,\,1\widehat{\sigma}_{\rm{VBF+Vh}}$& $\rho$ \\ \hline 
$\gamma\gamma$ & $1.32 $ & 0.8 & 0.38 & 0.7 & -0.30\\ \hline 
$ZZ^*$ & $1.70 $ & 0.3 & 0.4 & 1.20 & -0.59\\ \hline 
$WW^*$ & $0.98 $ & 1.28 & 0.28 & 0.55 & -0.20\\ \hline
$\tau\tau$ & $2 $ & 1.24 & 1.50 & 0.59 & -0.42\\ \hline
$b\bar{b}$ & 1.11 & 0.92 & 0.65 & 0.38 & 0 \\
\end{tabular}
\label{tab:Higgs data}
\end{ruledtabular}
\end{table}

In order to study the influence of new free parameters and to understand their correlations, 
we employ  the  minimum $\chi$-square method with the experimental data  considered.
For a given Higgs decay channel $f=\gamma\gamma, W W^*, Z Z^*, \tau \tau$, we define the $\chi^2_f$ as:
\begin{equation}
\chi^2_f = \frac{1}{\hat{\sigma}^2_{1}(1-\rho^2)}(\mu^{f}_{1} - \hat{\mu}^{f}_{1})^2
+ \frac{1}{\hat{\sigma}^2_{1}(1-\rho^2)}(\mu^{f}_{2} - \hat{\mu}^{f}_{2})^2 - \frac{2\rho}{\hat{\sigma}_{1}\hat{\sigma}_{2}(1-\rho^2)}(\mu^{f}_{1} - \hat{\mu}^{f}_{1})(\mu^{f}_{2} - \hat{\mu}^{f}_{2})\,, \label{eq:chi2f}
\end{equation}
where $\hat{\mu}^{f}_{1(2)}$, $\hat{\sigma}_{1(2)}$, and $\rho$ are the measured
Higgs signal strength, the one-sigma errors, and the correlation,
respectively.  The corresponding values are shown in 
Table~\ref{tab:Higgs data}. The indices $1$ and $2$ respectively stand for $\rm ggF+tth$ and $\rm VBF+Vh$, and 
$\mu^{f}_{1,2}$  are the results in the THDM. The global $\chi$-square  is defined by
\begin{equation}
\chi^2 = \sum_{f}\chi^2_{f} + \chi^2_{ST} \,,  \label{eq:chi2}
\end{equation}
where $\chi^2_{ST}$ is  the $\chi^2$ for S and T parameters;  its definition
can be obtained from Eq.(\ref{eq:chi2f}) by using the replacements $\mu^f_1 \to S^{\rm THDM}$ and $\mu^f_2\to T^{\rm THDM}$, and the corresponding values can be determined from Eq.~(\ref{test:ST}). 

Besides the bounds from theoretical considerations,  Higgs data, and upper limit $BR(\mu\to 3 e)< 1.0\times 10^{-12}$,  the schemes for the setting of parameters in this study are as follows:  the masses of SM Higgs and charged Higgs are fixed to be $m_h=125.5$ GeV and $m_{H^\pm}=500$ GeV, respectively, and  the regions of other involved parameters are chosen  as: 
\begin{align}
&\quad m_{H,A} \supset [126, 1000] \,{\rm GeV}\,,  \quad m^2_{12} \supset [-1.0, 1.5]\times 10^5\,{\rm GeV^2}\,,   \non \\
& \tan\beta \supset [0.5 , 50]\,,  \quad \alpha = [-\pi/2, \pi/2]\,.
\label{eq:inputs}
\end{align}
Since our purpose is to show the impacts of THDM on LFV, to lower the influence of the quark sector, we set ${\bf X}^q \sim 0$  in the current analysis; i.e., the Yukawa couplings of quarks behave like the type II THDM. The influence of $X^q \neq 0$ can be found elsewhere~\cite{Arhrib:2015maa}. To understand the small lepton FCNCs, we use the ansatz  $X^{\ell}_{ij} = \sqrt{m_i m_j}/v \chi^\ell_{ij}$; thus, $\chi^\ell_{ij}$ can be on the order of one. Although $h$-$\ell^+$-$\ell^-$ couplings also contribute to the $h\to 2\gamma$ process, unless one makes an extreme tuning on $\chi^\ell_{ii}$, their contributions to $h\to 2\gamma$ are small in the THDM. 

\begin{figure}[!ht] 
\includegraphics[width=0.55\textwidth]{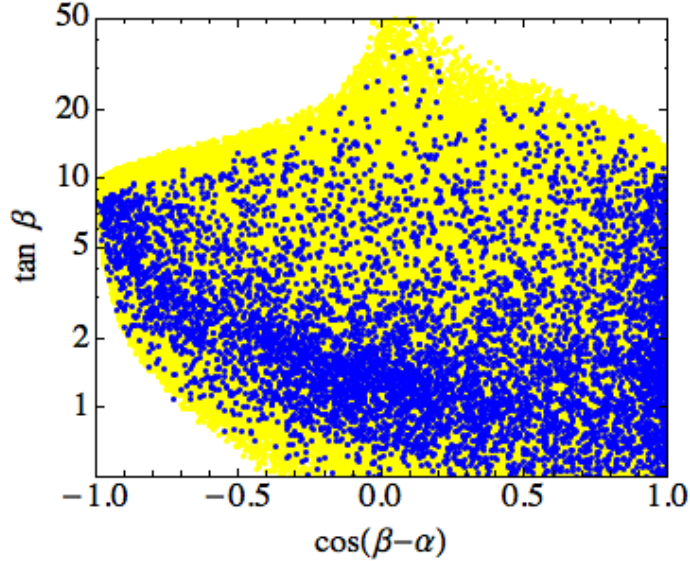} 
\caption{ Constraints from theoretical requirements and precision measurement of $\rho$-parameter (yellow dots) and  results (blue points)  when measurements of oblique parameters are included.}
\label{fig:C1}
\end{figure}

We now present the numerical analysis. Combining the theoretical requirements and $\delta\rho = (4.0 \pm 2.4)\times 10^{-4}$, the allowed ranges of $\tan\beta$ and $c_{\beta\alpha}$ are shown by the yellow dots in  Fig.~\ref{fig:C1} , where the scanned regions of Eq.~(\ref{eq:inputs}) were used. When the measurements of  oblique parameters are included, the allowed parameter space is changed slightly, as shown by blue dots in Fig.~\ref{fig:C1}. In both cases, data with $2\sigma$ errors are adopted. From the results, we see that the constraint on $c_{\beta\alpha}$ is loose  and the favorable range for $\tan\beta$ is $\tan\beta < 20$.  

\begin{figure}[!ht] 
\includegraphics[width=0.4\textwidth]{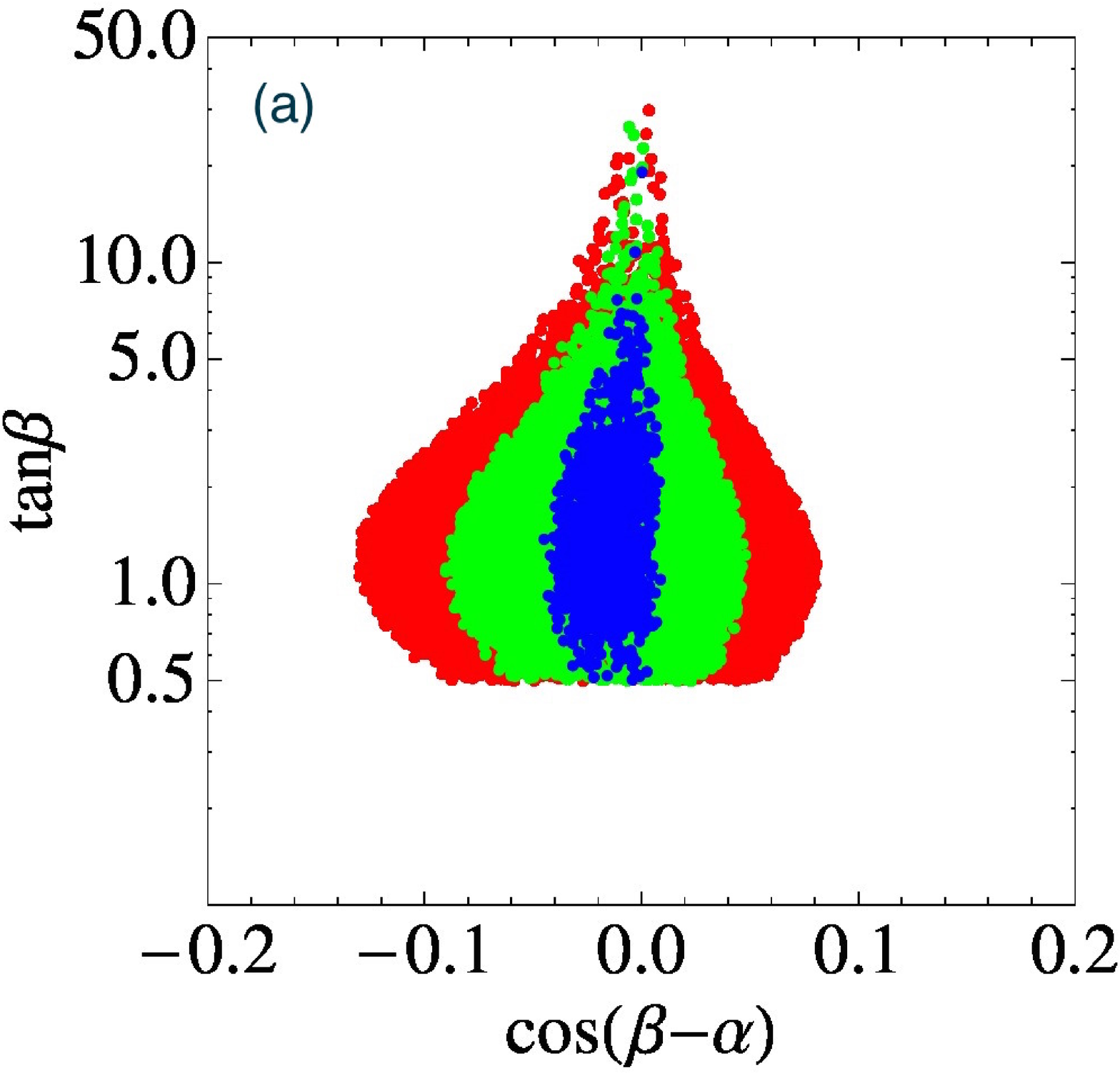} 
\includegraphics[width=0.4\textwidth]{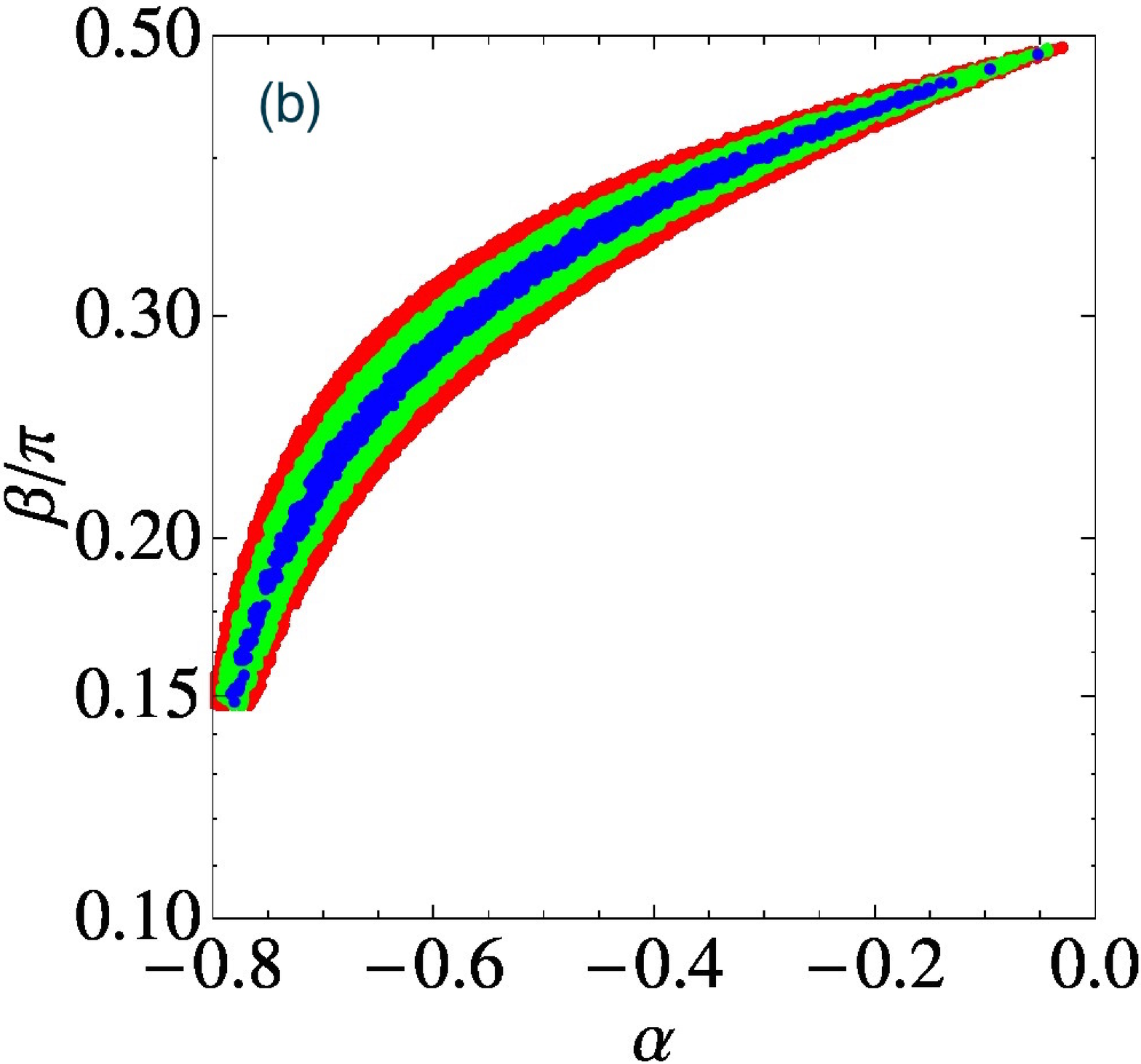}
\caption{ Bounds with $\chi$-square fit as a function of (a) $\tan\beta$ and $\cos(\beta-\alpha)$ and (b) $\beta/\pi$ and $\alpha$, where blue, green, and red denote $\Delta \chi^2 \leq 2.3$, $5.99$ ,and $11.8$, respectively.   }
\label{fig:C2}
\end{figure}

\begin{figure}[!ht] 
\includegraphics[width=0.4\textwidth]{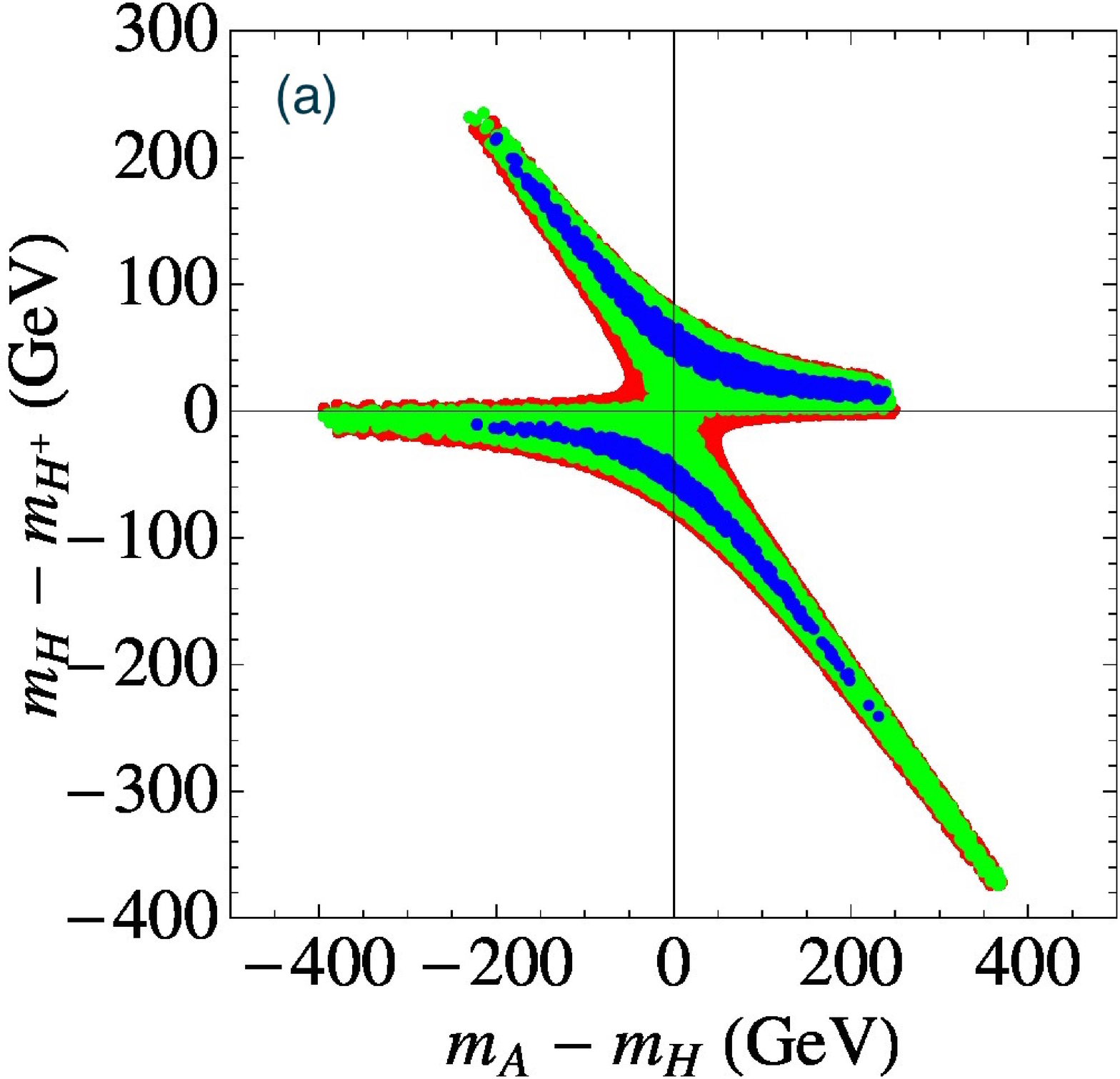} 
\includegraphics[width=0.4\textwidth]{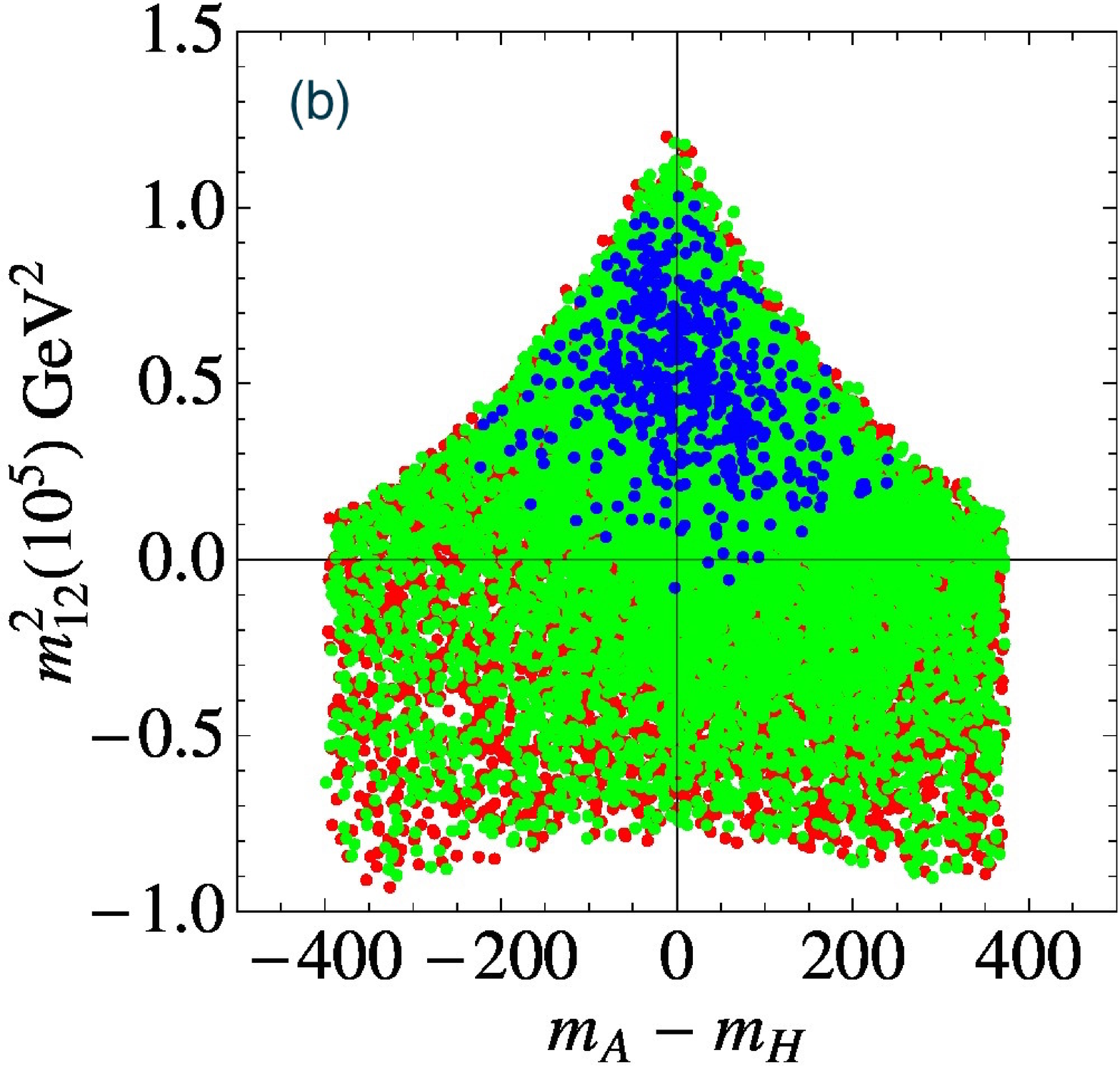}
\caption{   Correlations between (a) $m_H-m_{H^\pm}$ and $m_A-m_H$ and (b) $m^2_{12}$ and $m_A-m_H$,  where  blue, green, and red denote $\Delta \chi^2 \leq 2.3$, $5.99$, and $11.8$, respectively.   }
\label{fig:C3}
\end{figure}

To perform the constraints  from Higgs data listed in Table~\ref{tab:Higgs data},  we use the minimum $\chi$-square approach. The best fit is taken   at $68\%$, $95.5\%$, and $99.7\%$ CLs; that is, the corresponding errors of $\chi^2$ are $\Delta \chi^2 \leq 2.3$, $5.99$, and $11.8$, respectively. With the definitions in Eqs.~(\ref{eq:chi2f}) and (\ref{eq:chi2}), we present the allowed values of parameters  in Fig.~\ref{fig:C2}(a), where the  theoretical requirements, $\delta \rho$, oblique parameters, and Higgs data are all included. In the plots, blue, green, and red represent $68\%$, $95.5\%$, and $99.7\%$  CLs, respectively. It is clear that $c_{\beta\alpha}$ has been limited to a narrow range and that the favorable values of $\tan\beta$ are less than $10$. We can understand the correlation between angle $\beta$ and $\alpha$ from Fig.~\ref{fig:C2}(b). We will use these results to study other rare decays.  
For calculating  $\Delta a_\mu$ and  rare tau, $\mu$, and $Z$ decays, we need information about the allowed masses of $H$ and $A$. Using the results of $\chi$-square fitting, we present the correlation between $m_H-m_{H^\pm}$ and $m_A-m_H$ in Fig.~\ref{fig:C3}(a) and that  between $m^2_{12}$ and $m_A-m_H$ in Fig.~\ref{fig:C3}(b), where the ranges of parameters in Eq.~(\ref{eq:inputs}) are satisfied. 

\begin{figure}[hptb] 
\includegraphics[width=0.4\textwidth]{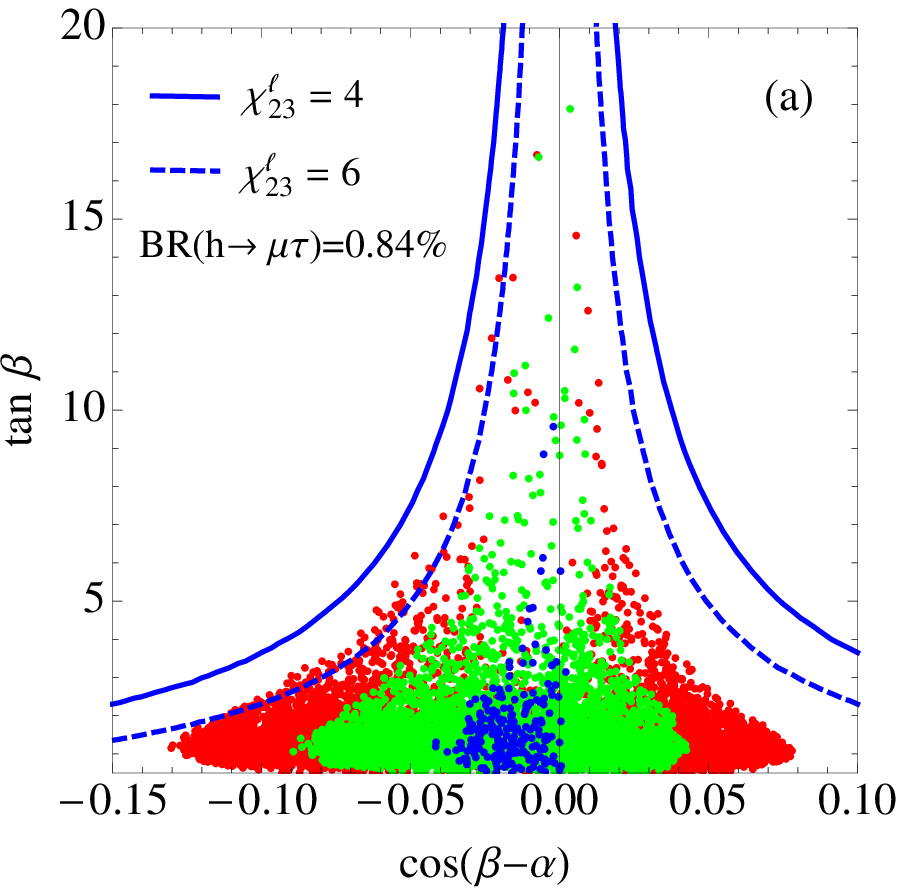} 
\includegraphics[width=0.43\textwidth]{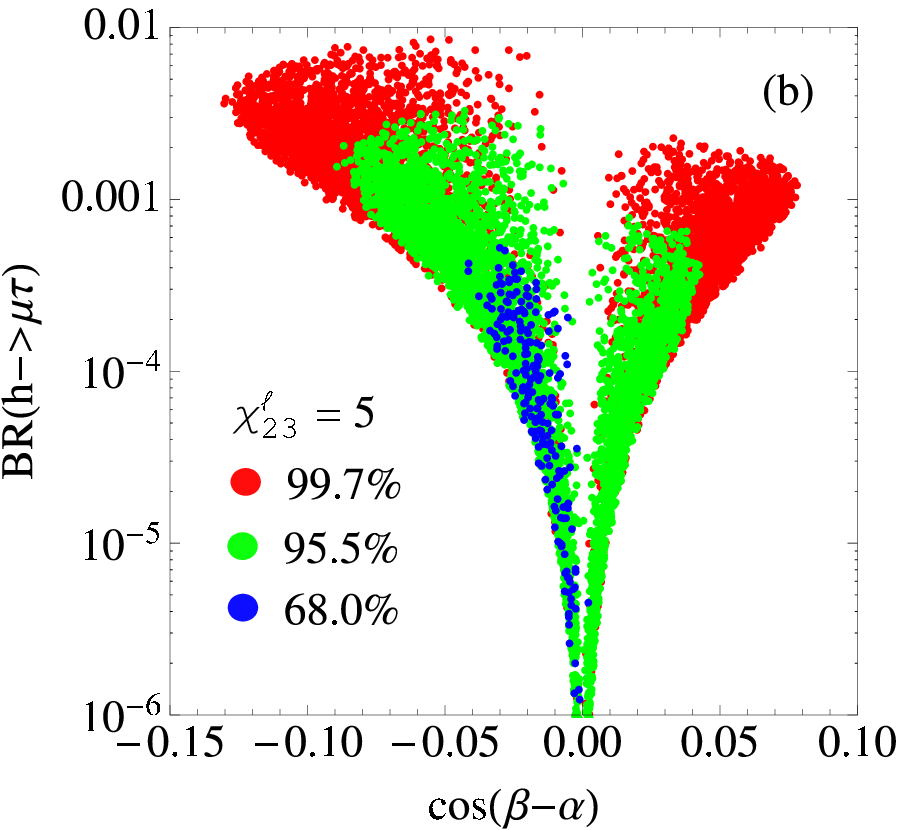}
\caption{  (a) Contour for $BR(h\to \mu \tau)=0.84\%$ as function of $\cos(\beta-\alpha)$ and $\tan\beta$ with $\chi^\ell_{23}=4$ (solid) and $6$ (dashed). (b) $BR(h\to \mu \tau)$ as  function of $\cos(\beta-\alpha)$, where blue, green, and red stand for the best fits at $68\%$, $95\%$, and $99.7\%$ CLs, respectively. }
\label{fig:brhmutau}
\end{figure}

After obtaining the allowed ranges of parameters, we analyze the implications of lepton-flavor-violating effects on $h\to \mu \tau$ and other rare decays.  From Eq.~(\ref{eq:brhmutau}), it can be seen that the $h\to \mu \tau$ decay is sensitive to $c_{\beta\alpha}$, $\tan\beta$, and $\chi^\ell_{23}$. In order to understand under what conditions the CMS results of $h\to \mu \tau$ can be reached in the type III THDM, we show the contour for $BR(h\to \mu\tau)=0.84\%$ as a function of $\tan\beta$ and $c_{\beta\alpha}$ in Fig.~\ref{fig:brhmutau}(a), where the solid and dashed lines stand for  $\chi^\ell_{23}=4$ and $6$, respectively. We find that in order to fit the central value of the CMS results and satisfy the bounds from Higgs data simultaneously, one needs $\chi^\ell_{23} > 5$. That is, with the severe limits of $\tan\beta$ and $c_{\beta\alpha}$, an accurate measurement of $h\to \mu \tau$ can directly bound the $\chi^\ell_{23}$. To clearly show the correlation between $BR(h\to \mu \tau)$ and the  parameters constrained by Higgs data, we plot the $BR(h\to \mu \tau)$ in terms of the results of Fig.~\ref{fig:C2} in Fig.~\ref{fig:brhmutau}(b), where we fix $\chi^{\ell}_{23}=5$ and blue, green, and red stand for the best fits at 68\%, 95\%, and 99.7\% CLs, respectively.

\begin{figure}[hptb] 
\includegraphics[width=0.41\textwidth]{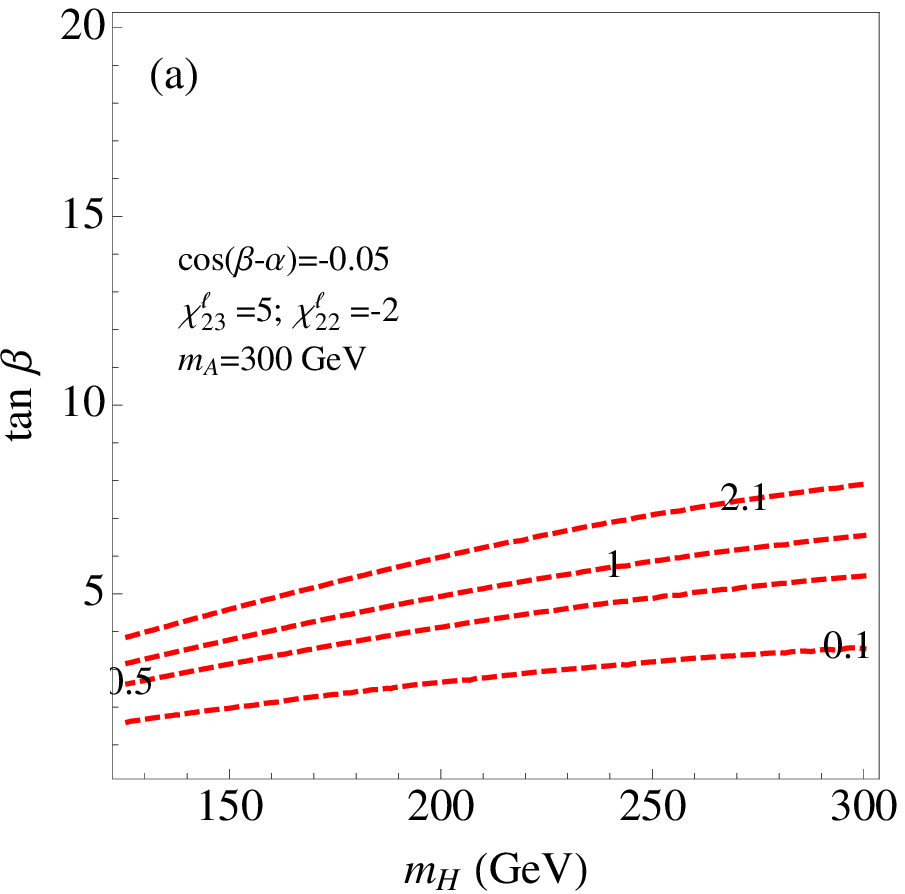} 
\includegraphics[width=0.4\textwidth]{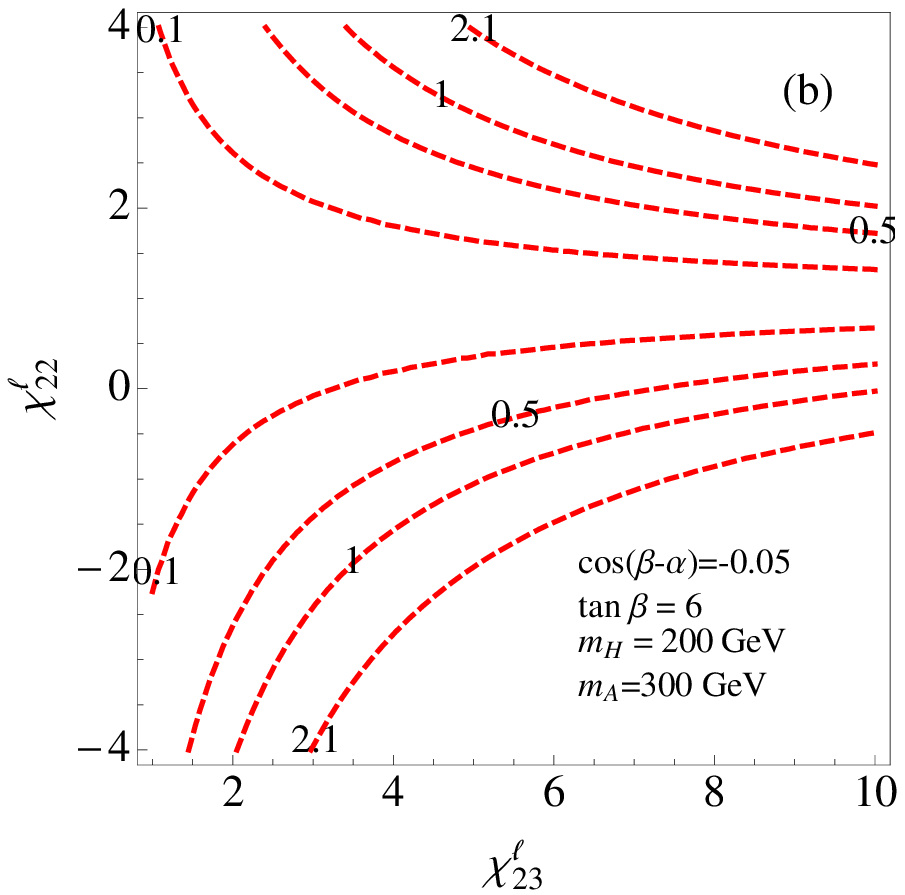}
\caption{  Contours for $BR(\tau\to 3\mu)\times 10^{8}$ as  function of (a) $m_H$ and $\tan\beta$ with $\chi^{\ell}_{23(22)}= 5 (-2)$ and (b) $\chi^{\ell}_{23}$ and $\chi^{\ell}_{22}$ with $m_H=200$ GeV and $\tan\beta=6$. In both plots,  $m_A=300$ GeV and $\cos(\beta-\alpha)=-0.05$.  }
\label{fig:brtau3mu}
\end{figure}

 From Eq.~(\ref{eq:tau3mu}), we see that the tree-level $\tau\to 3\mu$ decay is sensitive to the masses of $m_{H, A}$, $\tan\beta$, and $\chi^\ell_{23,22}$, but insensitive to  $c_{\beta\alpha}$. In Fig.~\ref{fig:brtau3mu}(a), we show the contours for $BR(\tau\to 3\mu) \times 10^{8}$ as a function of $\tan\beta$ and $m_H$, where $m_A=300$ GeV, $\chi^\ell_{23}=5$, $\chi^{\ell}_{22}=-2$, and $c_{\beta\alpha}=-0.05$ are used. The values in the plot denote the BR for $\tau\to 3\mu$;  the largest one is the current upper limit. Although a vanished $\chi^\ell_{22}$ still leads to a sizable $BR(\tau\to 3\mu)$,  its value influences the BR for the $\tau\to 3\mu$ decay. To understand the effect of $\chi^{\ell}_{22}$, we plot $BR(\tau\to 3\mu)\times 10^{8}$ as a function of $\chi^{\ell}_{23}$ and $\chi^{\ell}_{22}$ in Fig.~\ref{fig:brtau3mu}(b), where $\tan\beta=6$ and $m_{H(A)}=200 (300)$ GeV.  These parameter values are consistent with the constraints from Higgs data. 

\begin{figure}[!ht] 
\includegraphics[width=0.41\textwidth]{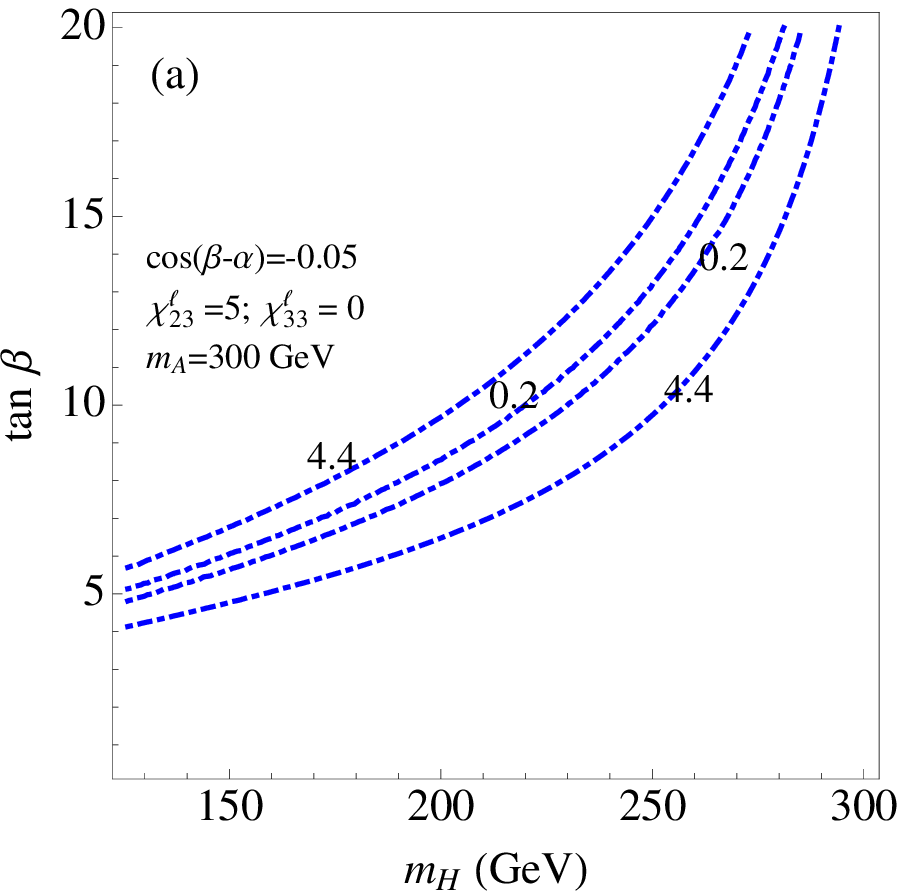} 
\includegraphics[width=0.4\textwidth]{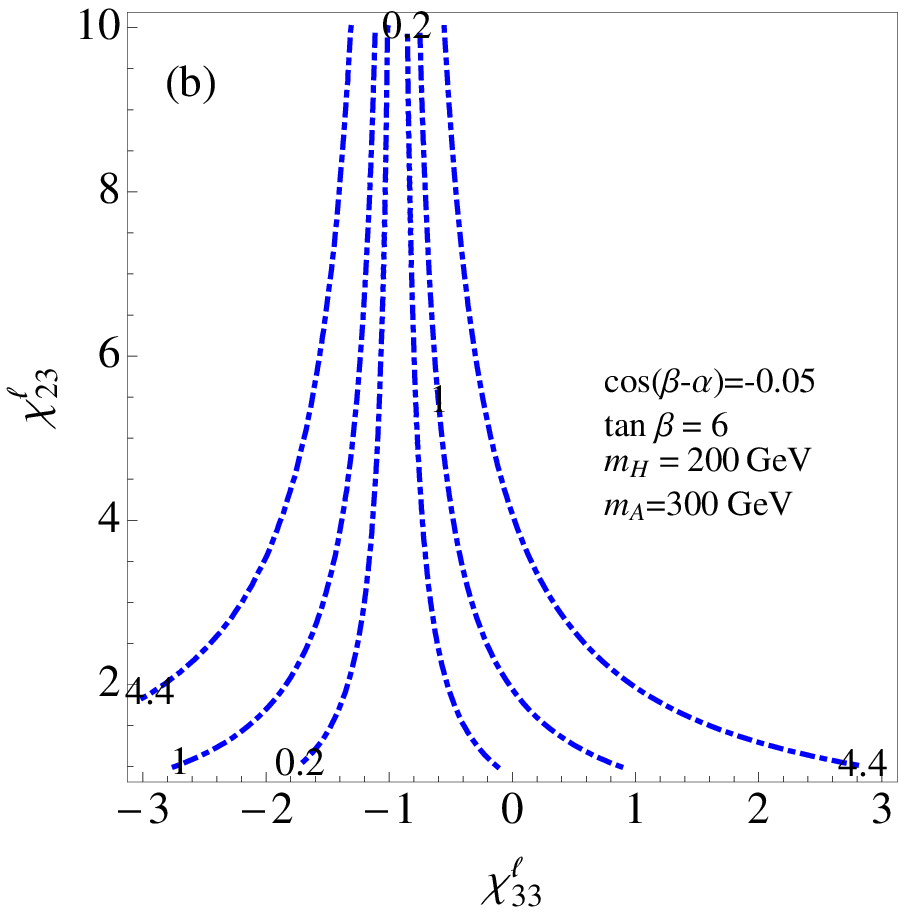} 
\caption{ Contours for $BR(\tau\to \mu \gamma)\times 10^{8}$  as  function of (a) $\tan\beta$ and $m_H$ with $\chi^\ell_{23}(\chi^\ell_{33})=5(0)$ and (b) $\chi^{\ell}_{23}$ and $\chi^{\ell}_{33}$ with $\tan\beta=6$ and $m_{H(A)}=200 (300)$ GeV.  One- and two-loop effects are included.  }
\label{fig:brtaumug}
\end{figure}

 From Eq.~(\ref{eq:WCps}), it can be seen that besides the parameters $\tan\beta$, $m_{H,A}$ and $\chi^{\ell}_{23}$,  $\tau\to \mu \gamma$ at the one-loop level is also dictated by $\chi^{\ell}_{33}$. Since $c_{\beta\alpha}$ has been limited to a narrow region, like the $\tau\to 3\mu$ decay, $\tau\to \mu \gamma$ is insensitive to $c_{\beta \alpha}$. We present the contours for $BR(\tau\to \mu\gamma)\times 10^{8}$ as a function of $\tan\beta$ and $m_H$ in Fig.~\ref{fig:brtaumug}(a), where we have included the one- and two-loop contributions and  $c_{\beta\alpha}=-0.05$, $\chi^\ell_{23(33)}=5 (0)$, and $m_A=300$ GeV. The largest value on the curves is the current experimental upper limit. We see that with strict constraints of Higgs data, $BR(\tau\to \mu \gamma)$ in the typeIII THDM  can still be compatible with the current upper limit when the decay $h\to \mu \tau$ matches CMS  observation.   

\begin{figure}[!ht] 
\includegraphics[width=0.41\textwidth]{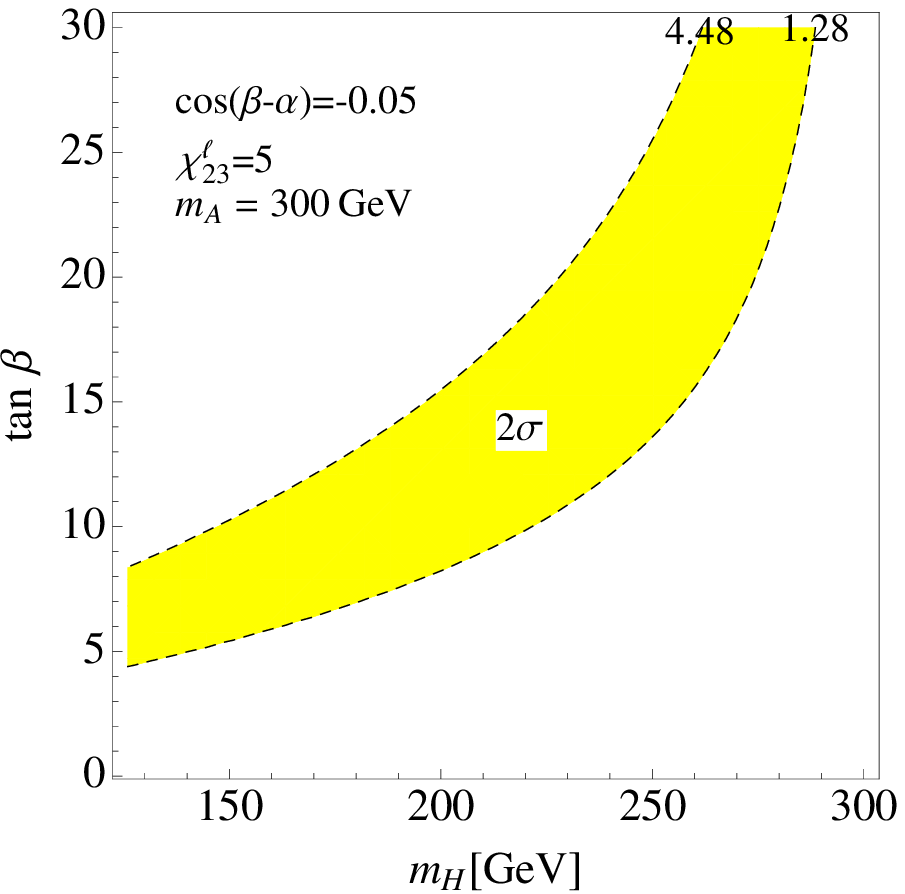} 
\includegraphics[width=0.4\textwidth]{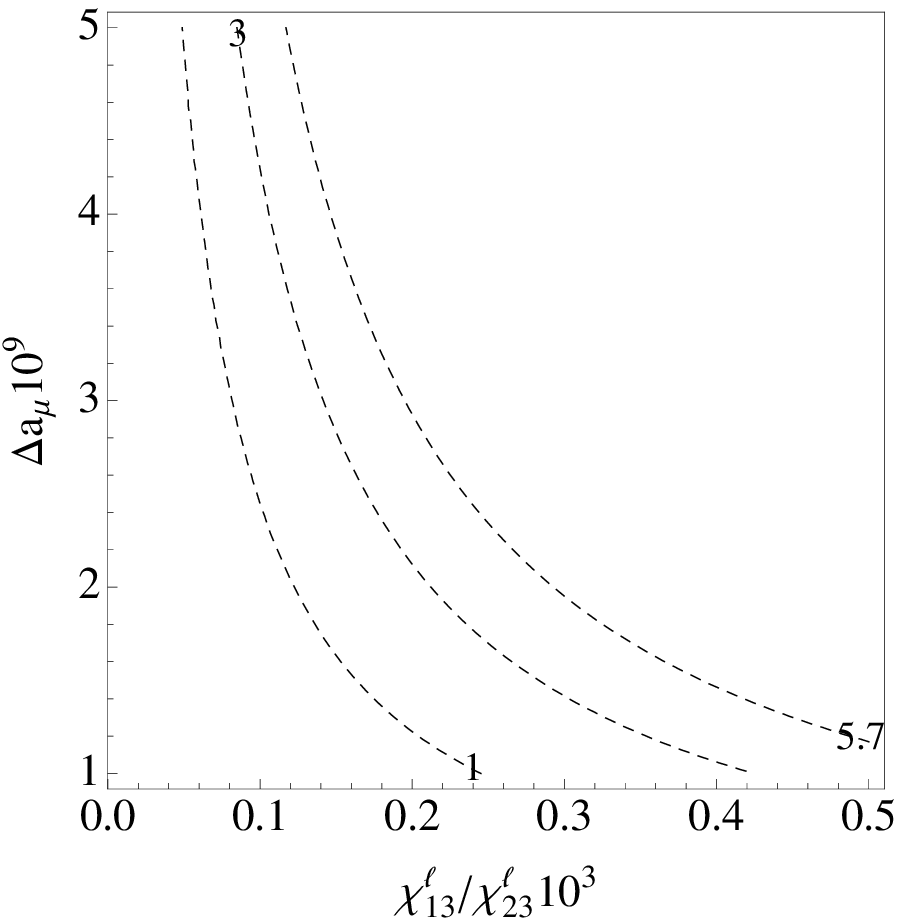} 
\caption{ (a) Contours for $\Delta a_\mu \times 10^{9}$  as  function of  $\tan\beta$ and $m_H$ with $m_A=300$ GeV, $\chi^\ell_{23}=5$, and $\cos(\beta-\alpha)=-0.05$ and (b) contours for $BR(\mu\to e \gamma)\times 10^{13}$ as a function of $\Delta a_\mu$ and $\chi^{\ell}_{13}/\chi^{\ell}_{23}$, where relation in Eq.~(\ref{eq:CphiL}) is adopted.   }
\label{fig:Damu}
\end{figure}

According to Eq.~(\ref{eq:g-2}), we know that muon $g-2$ strongly depends on $\chi^\ell_{23}$, $\tan\beta$, and $m_{H,A}$. It is of interest to determine whether $\Delta a_\mu$ could be explained by the type III model when the severe limits of involved parameters are imposed.  With $m_A=300$ GeV, $\chi^{\ell}_{23}=5$, we plot the contours for $\Delta a_\mu \times 10^{9}$ as a function of $\tan\beta$ and $m_H$ in Fig.~\ref{fig:Damu}(a), where the shaded region (yellow) stands for the central value with $2\sigma$ errors. From the plot, it is clear that these parameter values, which satisfy the Higgs data and $BR(h\to \mu \tau)=0.84\%$ can also make $(g-2)_\mu$  consistent with the discrepancy between the experimental data and SM prediction. Based on Eq.~(\ref{eq:CphiL}), it is found  that $\mu\to e \gamma$ can be expressed by $\Delta a_\mu$. With the ansatz $X^\ell_{ij}=\sqrt{m_i m_j}/v \chi^\ell_{ij}$, we show the contours for $BR(\mu\to e \gamma)$ as a function of $\Delta a_\mu$ and $\chi^{\ell}_{13}/\chi^\ell_{23}$ in Fig.~\ref{fig:Damu}(b), where the numbers on the curves are the BR for $\mu\to e \gamma$ decay obtained by multiplying  $10^{13}$. Clearly, in order to satisfy the bound from the rare $\mu \to e \gamma$ decay, $\chi^{\ell}_{13}$ has to be less than ${\cal O}(10^{-3})$. As a result, we get:
 \be
 BR(h\to e \tau) < 2 \times 10^{-4} \left( \frac{\chi^{\ell}_{13}/ \chi^\ell_{23}}{10^{-3}}\right)^2 BR(h\to \mu \tau).
 \ed
  Hence, in the type III THDM,  $h\to e \tau$ at least is an order of $10^{4}$  smaller than $h\to \mu \tau$.   

\begin{figure}[!ht] 
\includegraphics[width=0.41\textwidth]{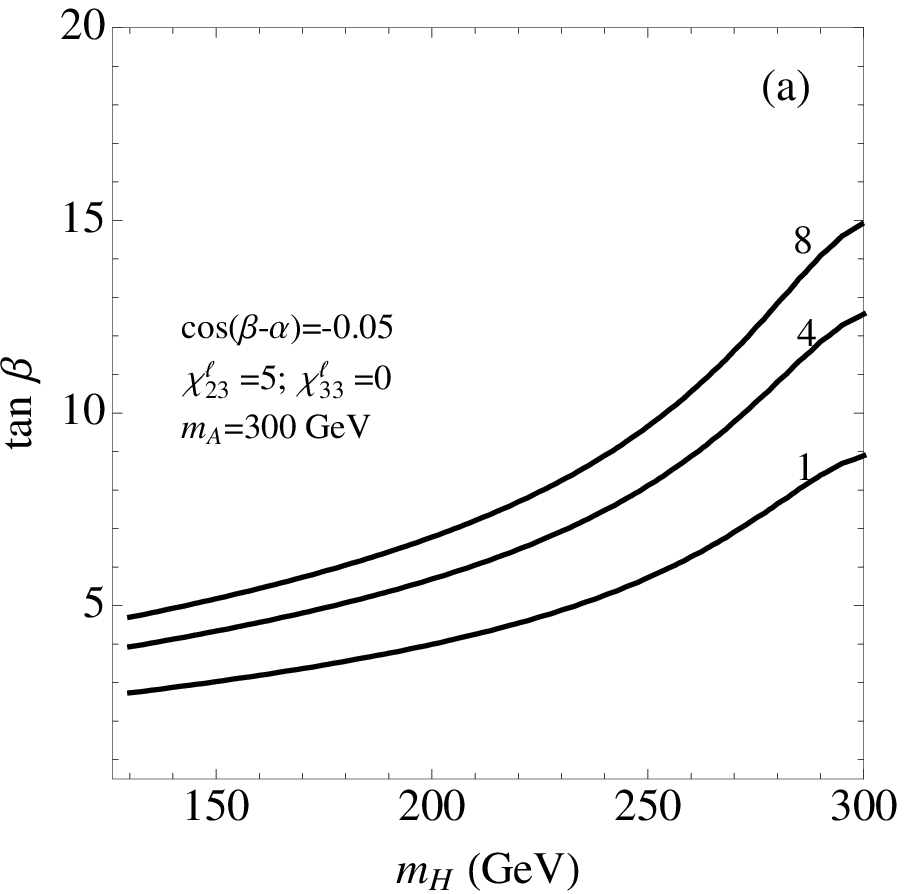} 
\includegraphics[width=0.4\textwidth]{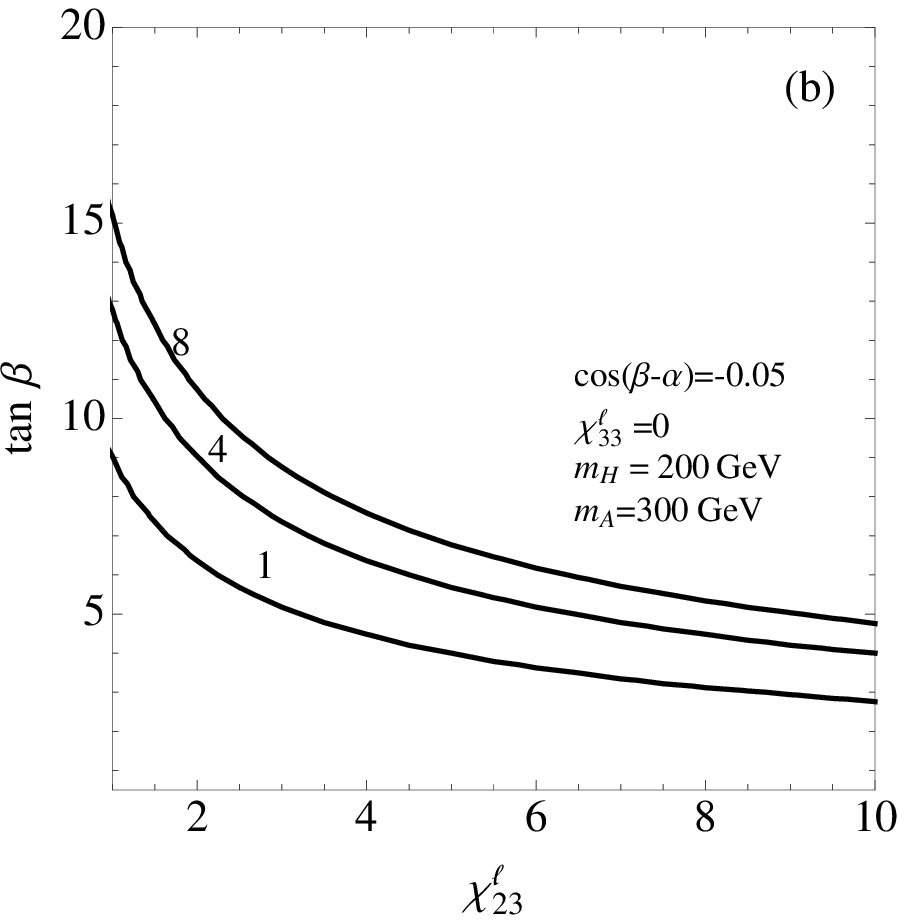} 
\caption{ Contours for $BR(Z\to \mu \tau) \times 10^{7}$  as  function of  (a) $\tan\beta$ and $m_H$ with $\chi^\ell_{23}=5$ and (b) $\tan\beta$ and $\chi^\ell_{23}$ with $m_H=200$ GeV. In both plots, we adopt $m_A=300$ GeV, $\chi^\ell_{33}=0$, and $\cos(\beta-\alpha)=-0.05$.   }
\label{fig:Zmutau}
\end{figure}

Finally, we discuss the decay $Z\to \mu \tau$. Similar to rare $\tau$ decays, $BR(Z\to \mu \tau)$ is sensitive to $\tan\beta$, $m_{H,A}$, and $\chi^{\ell}_{23(33)}$ in the type III model. Although we do not explicitly show the formulas in this paper, we  present the contours for $BR(Z\to \mu \tau)\times 10^{7}$ as a function of $\tan\beta$ and $m_H$ in Fig.~\ref{fig:Zmutau}(a), where $m_A=300$ GeV, $\chi^{\ell}_{23(33)}=5(0)$, and $c_{\beta\alpha}=-0.05$ are used. With the constrained parameters that fit the CMS results of $h\to \mu \tau$,  we find that BR for $Z\to \mu \tau$ decay is $BR(Z\to \mu \tau)< 10^{-6}$. The current experimental upper limit is $BR(Z\to \mu \tau)^{\rm exp} < 2.1\times 10^{-5}$. To understand the dependence of $\chi^{\ell}_{23}$,   we also show the contours as a function of $\tan\beta$ and $\chi^\ell_{23}$ with $m_H=200$ GeV in Fig.~\ref{fig:Zmutau}(b).

In summary, we  revisited the constraints for THDM. The bounds from theoretical requirements, precision $\delta \rho$, and oblique parameter measurements are shown in Fig.~\ref{fig:C1} and the bounds from Higgs data with $\chi$-square fit at $68\%$, $95.5\%$, and $99.7\%$ CLs are given in Fig.~\ref{fig:C2}. We clearly show the 
tension of Higgs data on the parameters of new physics. With the parameter values constrained by Higgs data, we find that the type III THDM can fit the CMS result $BR(h\to \mu\tau)=(0.84 ^{+0.39}_{-0.37})\%$. With the same set of parameters, the resultant branching ratios of tree-level $\tau\to 3\mu$ and loop-induced $\tau\to \mu \gamma$ decays are consistent with the current experimental upper limits. Under the strict limits of Higgs data, we clearly show that the anomaly of the muon anomalous magnetic moment can be explained by the type III model.  The rare decay $\mu\to e \gamma$ can be satisfied by small parameter $\chi^{\ell}_{13}$. As a result, we expect that the branching ratio for $h\to e \tau$ is smaller  than that for the decay  $h\to \mu \tau $ by  an order of magnitude of $10^{4}$. Additionally, we also calculated the branching ratio for rare decay $Z\to \mu \tau $ and the result is one order of magnitude smaller than the current experimental upper limit. \\

\noindent{\bf Acknowledgments}

The work of CHC was  supported by the Ministry of Science and Technology of  Taiwan,
R.O.C., under grant  MOST-103-2112-M-006-004-MY3.  The work of RB was supported by the Moroccan Ministry of Higher Education and Scientific Research MESRSFC and  CNRST: "Projet dans les
domaines prioritaires de la recherche scientifique et du d\'eveloppement
technologique": PPR/2015/6. \\

\appendix
\section{}

\subsection{ Yukawa couplings}

The Higgs Yukawa couplings to fermions are expressed as:
\be
-{\cal L}^h_{Y} &=& \bar u_L \left[ \frac{c_\alpha}{v s_\beta} {\bf m_u} - \frac{c_{\beta\alpha}}{ s_\beta} {\bf X}^u \right] u_R h + \bar d_L \left[ -\frac{s_\alpha}{v c_\beta} {\bf m_d} + \frac{c_{\beta\alpha}}{c_\beta} {\bf X}^d \right] d_R h \non \\
&+& \bar \ell_L \left[ -\frac{s_\alpha}{v c_\beta} {\bf m_\ell} + \frac{c_{\beta\alpha}}{ c_\beta} {\bf X}^\ell \right] \ell_R h + h.c.\,,
\ed
where  $c_{\beta\alpha} = \cos(\beta-\alpha)$, $s_{\beta\alpha}= \sin(\beta-\alpha)$ and ${\bf X}^fs$ are defined in Eq.~(\ref{eq:Xs}).
%
 Similarly, the Yukawa couplings of scalars $H$ and $A$ are expressed as:
 \be
-{\cal L}^{H, A}_{Y} &=& \bar u_L \left[ \frac{s_\alpha}{v s_\beta} {\bf m_u} + \frac{s_{\beta\alpha}}{s_\beta} {\bf X}^u \right] u_R H 
 + \bar d_L \left[ \frac{c_\alpha}{v c_\beta} {\bf m_d} - \frac{s_{\beta\alpha}}{ c_\beta} {\bf X}^d \right] d_R H \non \\
&+& \bar \ell_L \left[ \frac{c_\alpha}{v c_\beta} {\bf m_\ell} - \frac{s_{\beta\alpha}}{ c_\beta} {\bf X}^\ell \right] \ell_R H 
+ i  \bar u_L \left[ - \frac{\cot\beta}{v } {\bf m_u} + \frac{{\bf X}^u}{ s_\beta}  \right] u_R A  \non \\
&+& i \bar d_L \left[ -\frac{\tan\beta}{v } {\bf m_d} + \frac{{\bf X}^d}{ c_\beta}  \right] d_R A
+ i \bar \ell_L \left[ -\frac{\tan\beta}{v } {\bf m_\ell} + \frac{{\bf X}^\ell }{c_\beta}  \right] \ell_R A + h.c.
\ed
The Yukawa couplings of charged Higgs to fermions are:
\be
-{\cal L}^{H^\pm}_Y &=&  \sqrt{2} \bar d_L V^\dagger _{CKM} \left[ - \frac{ \cot\beta}{v } {\bf m_u} + \frac{{\bf X}^u}{ s_\beta}  \right] u_R H^- \non \\ 
&+& \sqrt{2} \bar u_L V_{\rm CKM} \left[ - \frac{ \tan\beta}{v } {\bf m_d} + \frac{{\bf X}^d}{  c_\beta}  \right] d_R H^+  \non \\
&+&\sqrt{2} \bar \nu_L V_{\rm PMNS} \left[ -\frac{\tan\beta}{v } {\bf m_\ell} + \frac{{\bf X}^\ell }{ c_\beta}  \right] \ell_R H^+ + h.c.\,,
\ed
where CKM and PMNS stand for Cabibbo-Kobayashi-Maskawa and Pontecorvo-Maki-Nakagawa-Sakata matrices, respectively. 
Except the factor $\sqrt{2}$ and  CKM matrix, the Yukawa couplings of charged Higgs are the same as those of pseudoscalar $A$. 

\subsection{$\tau\to \mu \gamma$ decay}
The effective interaction for $\tau \to \mu \gamma$ is expressed by
\be
{\cal L}_{\tau \to \mu \gamma} =\frac{ e }{16\pi^2} m_\tau \bar \mu \sigma_{\mu \nu} \left( C'_L P_L + C'_R P_R\right) \tau  F^{\mu \nu}\,,
\ed
where  the Wilson coefficients $C'_L$ and $C'_R$ from the one-loop neutral and charged scalars  are formulated  as:
\begin{align}
C'_{L(R)}& = \sum_{\phi = h, H, A, H^\pm} C^{\prime \phi}_{L(R)} \,, \non \\
C^{\prime h}_L & =  \frac{ c_{\beta\alpha}X^\ell_{32} }{2m^2_h c_\beta}y^\ell_{h 33}\left( \ln \frac{m^2_{h} }{m^2_\tau} - \frac{4}{3}\right) \,, \  %
C^{\prime H}_L  =  \frac{ -s_{\beta\alpha}X^\ell_{32} }{2m^2_H c_\beta}y^\ell_{H 33}\left( \ln \frac{m^2_{H} }{m^2_\tau} - \frac{4}{3}\right) \,,\non \\
C^{\prime A}_L & = - \frac{X^\ell_{32} }{2m^2_A c_\beta}y^\ell_{A 33}  \left( \ln \frac{m^2_{A} }{m^2_\tau} - \frac{5}{3}\right) \,, \ %
C^{\prime H^\pm}_L = - \frac{1}{12 m^2_{H^\pm}}  \left(\frac{\sqrt{2} X^{\ell}_{32}}{c_\beta}\right)y^{\ell}_{H^\pm 33}\,, \label{eq:WCps}
\end{align}
$C^{\prime h,H, A}_{R} = C^{\prime h,H,A}_{L}$ and $C^{\prime H^\pm}_R=0$. In addition, the contributions from  two loops  are given by~\cite{Sierra:2014nqa, Chang:1993kw, Davidson:2010xv}
\begin{align}
C^{h t(b)}_{2 L} & = C^{h t(b)}_{2 R} = 2 \frac{c_{\beta \alpha} X_{32} y^{u(d)}_{h 33} }{c_\beta} \frac{N_c Q_f^2 \alpha}{\pi} \frac{1}{m_\tau m_{t(b)}} f \left( \frac{m_{t(b)}^2}{m_h^2} \right) \,, \non \\ 
C^{H t(b)}_{2 L} & = C^{H t(b)}_{2 R}= -2 \frac{s_{\beta \alpha} X_{32} y^{u(d)}_{H 33} }{c_\beta} \frac{N_c Q_f^2 \alpha}{\pi} \frac{1}{m_\tau m_{t(b)}} f \left( \frac{m_{t(b)}^2}{m_H^2} \right)\,, \non  \\ 
C^{A t(b)}_{2 L} & = C^{A t(b)}_{2 R}= -2 \frac{ X_{32} y^{u(d)}_{A 33} }{c_\beta} \frac{N_c Q_f^2 \alpha}{\pi} \frac{1}{m_\tau m_{t(b)}} f \left( \frac{m_{t(b)}^2}{m_A^2} \right)\,, \non  \\ 
C^{W}_{2 L} & =C^{W}_{2 R}=  \frac{s_{\beta \alpha} c_{\beta \alpha} X_{32}}{c_\beta} \frac{g \alpha}{2 \pi m_\tau m_W}  \Bigl[ 3 f \left( \frac{m_W^2}{m_H^2} \right) + \frac{23}{4} g \left( \frac{m_W^2}{m_H^2} \right) + \frac{3}{4} h \left( \frac{m_W^2}{m_H^2} \right)\nonumber \\  &  \hspace{5cm} + \frac{m_H^2}{2 m_W^2} \left( f \left( \frac{m_W^2}{m_H^2} \right)- g \left( \frac{m_W^2}{m_H^2} \right) \right) \Bigr]  - (m_H \rightarrow m_h) \,,
\end{align}
where the loop functions are
\begin{align}
f(z) &= \frac{z}{2} \int^1_0 dx \frac{(1-2x(1-x))}{x(1-x)-z} \ln \frac{x(1-x)}{z}\,, \non \\
g(z) &= \frac{z}{2} \int^1_0 dx \frac{1}{x(1-x)-z} \ln \frac{x(1-x)}{z}\,, \non \\
h(z) &= -\frac{z}{2} \int^1_0 dx \frac{1}{x(1-x)-z} \left[1 - \frac{z}{x(1-x)-z} \ln \frac{x(1-x)}{z} \right] \,.
\end{align}
The BR for $\tau\to \mu \gamma$ is expressed by
 \be
 \frac{BR(\tau \to \mu \gamma)}{BR(\tau \to \mu \bar \nu_\mu \nu_\tau) } = \frac{3 \alpha_e}{4\pi G^2_F} \left( |C'_L|^2 + |C'_R|^2\right)\,.
 \label{eq:brtaumug}
 \ed

\end{document}